\documentclass{emulateapj}

\usepackage{epsfig,graphicx,rotating,portland,longtable}
\usepackage[indent,small,sc]{caption2}
\usepackage{alphalph,lscape}

\bibpunct{(}{)}{;}{a}{}{,}

\def\paren#1{\left( #1 \right)}

\shorttitle{Multiwavelength analysis of GRB~061126}
\shortauthors{Gomboc et al.}

\begin{document}

\title{Multiwavelength analysis of the intriguing GRB~061126: the reverse shock scenario and magnetization}

\author{A.~Gomboc\altaffilmark{1,2}, 
S.~Kobayashi\altaffilmark{2}, C.~Guidorzi\altaffilmark{2,3}, A.~Melandri\altaffilmark{2},  V.~Mangano\altaffilmark{4}, B.~Sbarufatti\altaffilmark{4}, 
C.~G.~Mundell\altaffilmark{2}, P.~Schady\altaffilmark{5}, R.~J.~Smith\altaffilmark{2}, A.~C.~Updike\altaffilmark{6}, D.~A.~Kann\altaffilmark{7}, 
K.~Misra\altaffilmark{8,9}, E.~Rol\altaffilmark{10}, A.~Pozanenko\altaffilmark{11}, A.~J.~Castro-Tirado\altaffilmark{12}, 
G.~C.~Anupama\altaffilmark{13}, 
D.~Bersier\altaffilmark{2}, M.~F.~Bode\altaffilmark{2},
D.~Carter\altaffilmark{2}, P.~Curran\altaffilmark{14}, A.~Fruchter\altaffilmark{15}, J.~Graham\altaffilmark{15}, D.~H.~Hartmann\altaffilmark{6}, M.~Ibrahimov\altaffilmark{16}, 
A.~Levan\altaffilmark{17}, A.~Monfardini\altaffilmark{2,18}, C.~J.~Mottram\altaffilmark{2},  P.~T.~O'Brien\altaffilmark{10}, P.~Prema\altaffilmark{19}, D.~K.~Sahu\altaffilmark{20}, I.~A.~Steele\altaffilmark{2}, 
N.~R.~Tanvir\altaffilmark{10}, K.~Wiersema\altaffilmark{10}}

\email{andreja.gomboc@fmf.uni-lj.si, (ag, sk, crg, axm, cgm, rjs, dfb, mfb, dxc, am, cjm, ias)@astro.livjm.ac.uk}
\altaffiltext{1}{Faculty of Mathematics and Physics, University of Ljubljana, Jadranska 19, SI-1000 Ljubljana, Slovenia.}
\altaffiltext{2}{Astrophysics Research Institute, Liverpool John Moores University, Twelve Quays House, Birkenhead, CH41 1LD, UK.}
\altaffiltext{3}{INAF - Osservatorio Astronomico di Brera, via Bianchi 46, 23807 Merate (LC), Italy.}
\altaffiltext{4}{INAF - Instituto di Astrofisica Spaziale e Fisica Cosmica di Palermo, via Ugo La Malfa 153, 90146 Palermo, Italy.}
\altaffiltext{5}{The UCL Mullard Space Science Laboratory, Holmbury St Mary, Dorking, Surrey RH5 6NT, UK.}
\altaffiltext{6}{Department of Physics and Astronomy, Clemson University, Clemson, SC 29634, USA.}
\altaffiltext{7}{Th\"{u}ringer Landessternwarte Tautenburg, Sternwarte 5, 07778 Tautenburg, Germany.}
\altaffiltext{8}{Aryabhatta Research Institute of Observational Sciences, Manora Peak, Nainital 263 129, India.}
\altaffiltext{9}{Inter University Center for Astronomy and Astrophysics (IUCAA), Post Bag 4, Ganeshkhind, Pune 411 007, India.}
\altaffiltext{10}{Department of Physics and Astronomy, University of Leicester, Leicester, LE1 7RH, UK.}
\altaffiltext{11}{Space Research Institute (IKI), 84/32 Profsoyuznaya Str, Moscow 117997, Russia.}
\altaffiltext{12}{Instituto de Astrof\'{\i}sica de Andaluc\'{\i}a (CSIC), P.O. Box 03004, E-18080 Granada, Spain.}
\altaffiltext{13}{Indian Institute of Astrophysics, Bangalore, 560 034, India.}
\altaffiltext{14}{Astronomical Institute Anton Pannekoek, University of Amsterdam, Kruislaan 403, 1098 SJ Amsterdam, The Netherlands.}
\altaffiltext{15}{Space Telescope Science Institute, 3700 San Martin Drive, Baltimore, MD 21218, USA.}
\altaffiltext{16}{Ulugh Beg Astronomical Institute, Tashkent 700052, Uzbekistan.}
\altaffiltext{17}{Department of Physics, University of Warwick, Coventry CV4 7AL, UK.}
\altaffiltext{18}{CNRS-CTBT, Grenoble 38000, France.}
\altaffiltext{19}{Institute of Astronomy, University of Cambridge, Madingley Road, Cambridge, CB3 0HA, UK.}
\altaffiltext{20}{Center for Research and Education in Science \& Technology, Hosakote, Bangalore, 562 114, India.}

\begin{abstract} 
We present a detailed study of the prompt and afterglow emission from {\em Swift} GRB~061126 using BAT, XRT, UVOT data and multi-color optical imaging from ten ground-based telescopes.
GRB~061126 was a long burst ($T_{90}=191$~s) with four overlapping peaks in its $\gamma$-ray light curve. The X-ray afterglow, observed from 26 min to 20 days after the burst, shows a simple
power-law decay with $\alpha_{\rm X}=1.290\pm0.008$. Optical observations presented here cover the time range from 258~s (Faulkes Telescope North) to 15~days (Gemini North) after the burst; the decay rate of the optical afterglow shows a steep-to-shallow transition (from $\alpha_1=1.48\pm 0.06$ to $\alpha_2=0.88\pm0.03$) approximately 13~min after the burst. 
We suggest the early, steep component is due to a reverse shock and show that the magnetic energy density in the ejecta, expressed as 
 a fraction of the equipartion value, is a few ten times larger than 
 in the forward shock in the early afterglow phase. The ejecta might be 
 endowed with primordial magnetic fields at the central engine.
The optical light curve implies a late-time break at about 1.5~days after the burst, while there is no evidence of the simultaneous break in the X-ray light curve.
We model the broad band emission and show that some afterglow characteristics (the steeper decay in X-ray and the shallow spectral index from optical to X-ray) are difficult to explain in the framework of the standard fireball model. This might imply that the X-ray afterglow is due to an additional emission process, such as late time central engine activity rather than blast-wave shock emission. The possible chromatic break at 1.5~days after the burst would give support to the additional emission scenario.

\end{abstract}


\section{Introduction} 

Facilitated by the rapid accurate localization and dissemination of
observed properties of Gamma Ray Bursts (GRBs) by the {\em Swift}
satellite \citep{geh04}, multiwavelength studies of GRBs are providing
important insights into the physics of these prodigious cosmic
explosions \citep{zha07, mes1, pir}. Despite the diverse range
of observed properties of GRBs when studied over a large time range in
wavebands spanning the electromagnetic spectrum, the primary goals of
multiwavelength analyses are to understand the physical origin of
prompt and afterglow emission, to challenge current theoretical
models, to determine the nature of the expanding fireball and the role
played by magnetic fields in driving the explosion.

The combination of $\gamma$-ray, X-ray, optical and ultraviolet data
from {\em Swift} instruments with deep, early-time optical
imaging from rapid-response ground-based robotic telescopes, such as
the Faulkes and Liverpool telescopes, as well as later-time
observations with 4- and 8-m class telescopes has provided
unprecedented datasets for the investigation of GRB physics.

Here we present a detailed analysis of a set of multiwavelength
observations of {\em Swift} GRB~061126 comprising $\gamma$-ray, X-ray,
ultraviolet and optical observations from ground- and space-based
telescopes that observed the initial prompt emission and early
afterglow through to the late stages of the fading afterglow, 15$-$20
days after the burst. Following the detection of GRB~061126 by {\em
Swift's} Burst Alert Telescope (BAT), several ground-based telescopes
(Raptor-S, Super-LOTIS, NMSU-1m telescope, PAIRITEL, Faulkes Telescope North, 
KAIT, and 0.3-m telescope at New Mexico Skies Observatory)
reacted promptly to the BAT trigger and detected a
bright optical and NIR afterglow, with detections being obtained in
the first tens to hundreds of seconds after the burst. The {\em Swift}
satellite did not slew immediately to the burst location because of
the Earth limb constraint. Therefore, observations with the narrow
field instruments, the X-Ray Telescope (XRT) and the UV/Optical
Telescope (UVOT), began 26 minutes after the trigger. An associated host galaxy
was detected \citep{rol,per} with a redshift $z=1.1588$ \citep{per}.

The optical afterglow of GRB~061126 shows a steep-to-flat transition
at $\sim$13~min after the trigger. Similar flattening has been
observed in optical afterglows of GRB~990123 \citep{ake}, GRB~021211
\citep{fox, pan, li}, GRB~060117 \citep{jel1}, and GRB~080319B \citep{rac}.  Particularly, the
GRB~021211 occurred at a similar redshift  
\citep[$z=1.006$,][]{vre} and
was linked with a possible supernova \citep{del}.  In these cases, the
early, steep afterglow was interpreted as due to emission from the
reverse shock dominating the light curve, while the later, more slowly
fading component as due to the forward shock \citep{sar12, nak, wei, fox,
pan, jel2}.  Some of these bursts have also shed light
on the issue of magnetization of the fireball. It is shown that at the 
deceleration of a fireball ejecta, the microscopic parameter $\epsilon_B$ 
in the ejecta should be much larger than in the forward shock in the 
case of GRB~990123 and possibly GRB~021211 \citep{zha, kum, fan1}.

With its steep-to-shallow optical light curve behavior, GRB~061126
offers a valuable opportunity to investigate the multiwavelength
prompt and afterglow properties of a GRB with a prominent reverse
shock component, which is not always present in light curves of GRBs
with bright optical counterparts \citep{mun07a}. Observations and data
reduction are presented in \S 2; the derived temporal and spectral
characteristics of the burst are presented in \S 3, and in \S 4 we
present and discuss a reverse and forward shock model, implications
for the standard model and the magnetization of the fireball.

Throughout this paper we use the following notation for a power-law
flux: $F(\rm \nu, t) \propto t^{-\alpha} \nu^{-\beta}$, where $\alpha$
is the temporal decay index, $\beta$ is the spectral index and it is
related to the photon index $\Gamma$ as $\Gamma=1+\beta$. Quoted
errors are given at $1 \sigma $ confidence level, unless stated
otherwise.

\section{Observations and Analysis}

\subsection{{\em Swift} BAT data}

BAT triggered and localized GRB~061126 (BAT trigger 240766) on 2006 Nov 26, at 08:47:56 UT \citep{sba1}. 
We refer to this time as $T_0$ throughout the paper.
This was a 1.024 s rate-trigger on a long burst with $T_{90}=191$~s. The BAT light curve in different energy bands is shown in Figure~\ref{fig:gammalc}.

BAT data were obtained in the burst mode, covering $T_0-239$~s to $T_0+574$~s \citep{kri} and 
 were processed using the HEASOFT software package, version 6.1.2 and
version 2.6 of the Calibration DataBase, applying calibration, standard
filtering and screening criteria.
We extracted the mask-tagged light curves (Figures~\ref{fig:gammalc} and \ref{fig:gammatail}) with a binning time of 64~ms in the four nominal energy bands adopting the ground-refined coordinates provided by the BAT team \citep{kri}.
 We applied the energy calibration using the closest-in-time gain/offset file through the tool {\tt bateconvert}. The light curves are expressed as count rates: these are background-subtracted counts per second per fully illuminated detector for an equivalent on-axis source, as the default corrections are
 applied: {\tt ndets, pcode, maskwt, flatfield}.
 We extracted the mask weighted spectrum from $T_0-10$~s to $T_0+200$~s  using the tool {\tt batbinevt}.
All required corrections were applied: we updated it through {\tt batupdatephakw} and generated the detector response matrix using {\tt batdrmgen}. We then used {\tt batphasyserr} to account for the BAT systematics as a function of energy.
Finally we grouped the energy channels of the spectrum by imposing a 5-$\sigma$ threshold on each grouped channel. 
The spectrum 
(Figure~\ref{fig:gammaour}) 
was fit with XSPEC11.3.

\subsection{{\em Swift} XRT data}

XRT began observing the burst at 09:14:31 UT, i.e. at $T_0+1598$~s, and 
monitored the source until 2006 December 28 at 23:59:57 UT for a total of 29 
observation sequences. 

XRT data were processed using the HEASOFT package.
The XRT exposure times after all the cleaning procedures were 203 s in Window 
Timing mode (WT) and 271 ks in Photon Counting mode (PC), distributed over a 
time interval of 32 days. PC data from the first sequence were corrected for pile-up, caused by the relatively 
high count rate of the source. The XRT light curve 
(Figure~\ref{fig:lc}) 
was extracted requiring a minimum signal-to-noise ratio of 3.

The spectral analysis was performed only on the first seven sequences (up to $T_0+570$~ks; the source dropped below count 
rate $3\times 10^{-3}$~counts~s$^{-1}$ afterwards). Data in three time intervals:
WT data from the first sequence (from $T_0+1603$~s to $T_0+1807$~s), PC data from the first sequence (from $T_0+1807$~s to $T_0+15280$~s) and PC data from sequences
 2-7 (from $T_0+15280$~s to $T_0+570$~ks), 
 were 
fitted with an absorbed power-law model using XSPEC version 11.3.
Instrumental energy channels below 0.3~keV and above 10~keV for PC and WT
spectra were ignored.  Data were binned with a requirement of a minimum of 20 photons 
per bin. 
Auxiliary response files and exposure maps were created 
using the HEASOFT software for each segment, and the appropriate response 
matrixes from the CALDB were applied.

\subsection{{\em Swift} UVOT data}
The UVOT began observing the field of the GRB~061126 at $T_0+1605$~s. Observations started with a 9~s settling 
exposure in $V$-filter, followed by a 100~s exposure in white light filter. After this the automated sequence rotated six times through the UVOT filters, taking a series of short exposures ($UVW1, U, B$, white, $UVW2, V, UVM2$; 10~s for white filters and 20~s for the rest). Observations continued with the rotating filter wheel and a combination of exposures of 200~s, 300~s or 900~s up to $T_0+50$~ks. Details of the UVOT observation log are in \citet{sba3}.

To improve the signal to noise ratio of the afterglow detection, consecutive images were coadded to create at least 40~s exposures. Photometric measurements were obtained from the UVOT data with the tool {\sc uvotsource} (version 2.2) using a circular source extraction region with a 
$3\arcsec$ and $4.5\arcsec$ radius for the optical and UV filters, respectively. An aperture correction was then applied to the photometry to maintain compatibility with the current UVOT effective area calibration\footnotemark[1]. The background was measured in a source-free region near the target using an extraction radius of $12\arcsec$. 

To combine UVOT data with ground based observations, we re-calibrated the UVOT B and V values with respect to the 5 field stars detected also in ground based B- and V-band observations. Due to similarity of calibration stars' colors the color correction between UVOT and standard filter magnitudes could not be applied. 

\footnotetext[1]{http://heasarc.gsfc.nasa.gov/docs/caldb/swift/docs/uvot/}

\subsection{Ground-based Optical data}
Observations with ground-based telescopes started shortly after the trigger time: Raptor-S at $T_0+20.87$~s \citep{wre}, Super-LOTIS at $T_0+35$~s \citep{wilmil}, NMSU-1m telescope at $T_0+47$~s \citep{hol}, PAIRITEL at $T_0+58$~s \citep{blo}, Faulkes Telescope North (FTN) at $T_0+258$~s \citep{smi}, KAIT at $T_0+356$~s \citep{per} and 0.3-m telescope at New Mexico Skies Observatory at $T_0+623$~s \citep{tor}.
The FTN  reacted robotically and using the automatic GRB-pipeline LT-TRAP \citep{guiTRAP}, independently
detected the fading optical afterglow at the position in
agreement with the position from UVOT: R.A.(J2000)=05:46:24.46, Dec(J2000)=$+$64:12:38.5 
($\pm 0.5\arcsec$) \citep{van}. Detection of the IR afterglow at a consistent position followed shortly by \citet{blo}. 

Observations continued with several telescopes, including the SARA telescope \citep{upd}, the Mt Abu IR Observatory (MIRO) \citep{bal}, the Tautenburg  Schmidt telescope (TLS) \citep{kan},
the Sampurnanand Telescope (ST) \citep{mis}, the Maidanak observatory (MAO) \citep{poz}, the Himalayan Chandra Telescope (HCT), and the Observatorio de Sierra Nevada (OSN) 1.5m telescope.
Late time observations ($> T_0+1$~day) were performed by the Liverpool Telescope (LT) as part of the 
{\it RoboNet-1.0} project\footnotemark[2] \citep{gom}, 
as well as with the TLS, Isaac Newton Telescope (INT) and Gemini North.
Details of the ground-based observations presented in this paper are summarized in Table \ref{observations}.

\footnotetext[2]{http://ww.astro.livjm.ac.uk/RoboNet/}

Magnitudes in $BVR$ bands have been calibrated using Landolt
standard field stars \citep{lan} observed by the FTN on the same night
as the GRB. The night was photometric and the zero point of each
optical filter was stable throughout the entire FTN observational
sequence. Photometry was performed using the Starlink GAIA
Photometry Tool, carefully selecting the right parameters for each
observation acquired with different instruments. Data taken by other
telescopes were then cross-calibrated with the FTN observations using
several stars in the field to provide a consistent and well-calibrated
multi-telescope light curve. Data from the LT, INT and Gemini North, as well as FTN $i'$-band observations, were
calibrated using the SDSS pre-burst (revised) photometry \citep{coo}.  Finally,
the data were corrected for the Galactic
extinction: $E_{B-V}=0.182$~mag derived from the extinction maps by
\citet{sch} and $A_V=0.604$~mag (following \citet{car} we evaluate
$A_B=0.79$~mag, $A_R=0.49$~mag, $A_{i'}=0.39$~mag and
$A_{g'}=0.70$~mag). Conversion from magnitudes to flux densities
 followed \citet{bes} and \citet{fuk}.

\section{Results}

\subsection{Prompt Gamma-ray Emission}

\begin{figure}
\centering
\includegraphics[angle=0, scale=0.7]{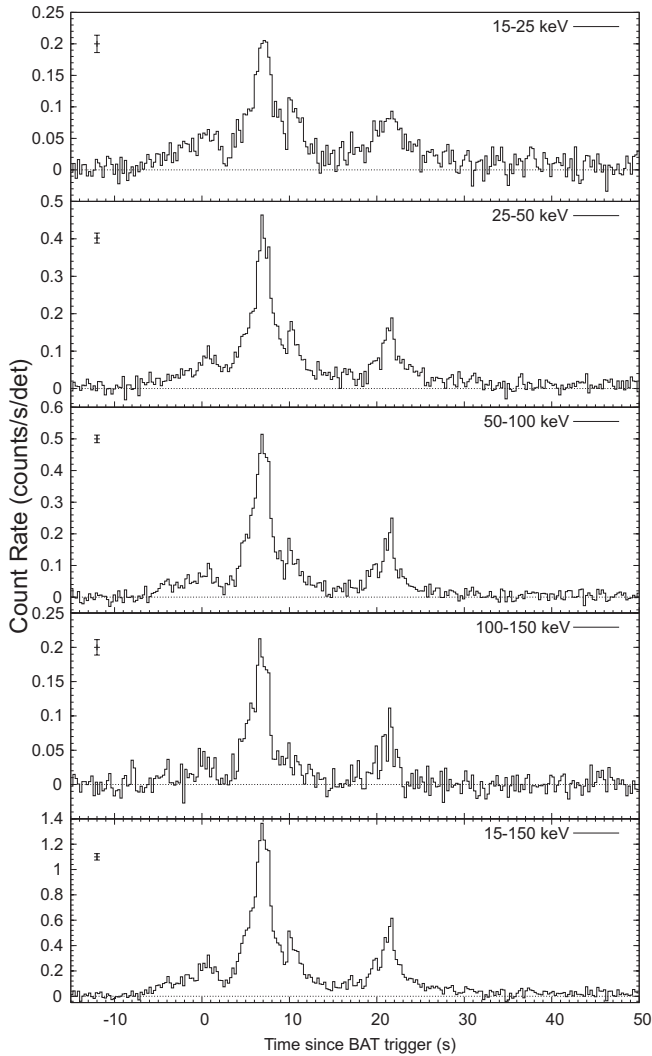}
\caption{From top to bottom: the BAT light curve of GRB~061126 during the main activity period in 15-25 keV, 25-50 keV, 50-100 keV, 100-150 keV energy bands and the sum (15-150 keV) in the bottom panel, respectively. Typical error bars are shown on the top left of each panel.}
\label{fig:gammalc}
\end{figure}

\begin{figure}
\centering
\includegraphics[angle=0, scale=0.7]{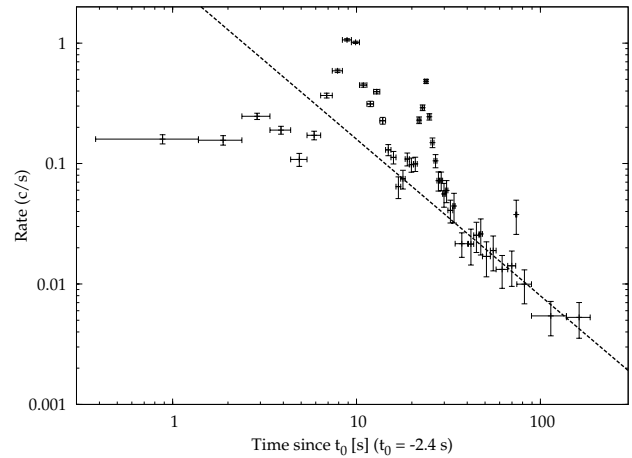}
\caption{The BAT 15-150~keV light curve with a logarithmic scale. After the second main peak the gamma tail is evident. The dashed line shows
the power-law fit with $\alpha_\gamma =1.3$.}
\label{fig:gammatail}
\end{figure}

The prompt emission of GRB~061126 shows in all BAT energy bands two main peaks and two smaller ones 
(Figure~\ref{fig:gammalc}).
The mask-weighted light curves show emission above background level starting at $T_0-10$~s. The brightest peak occurs at $T_0+7$~s and 
the last peak, which is also second brightest ends at $\approx T_0+25$~s. Low level emission is ongoing to $\approx T_0+200$~s resulting in $T_{90}(15 - 350$~keV$)=191\pm 10$~s. This $\gamma$-ray emission tail is more evident in the logarithmic scale (Figure~\ref{fig:gammatail}) and can be fitted with a power-law of the form $\propto (t-t_{s})^{-\alpha_\gamma}$. Using data points at $t>T_0+37$~s (after the last peak), we derive best fit parameters $\alpha_\gamma = 1.3\pm 0.2$, $t_s=(-2.4\pm 12.2)$~s with $\chi^2/{\rm dof}=6.9/9$.

\begin{figure}
\centering
\epsscale{1.0}
\includegraphics[angle=-90,scale=0.35]{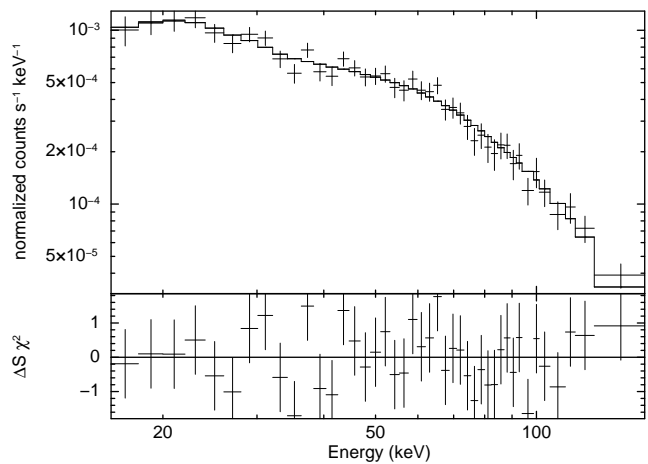}
\caption{The BAT total spectrum from $T_0- 10$~s to $T_0+200$~s and the fit with the Band function and parameters  $\alpha_{\rm B}$ = $-1.05\pm0.17$,
$\beta_{\rm B}$= -2.3 (fixed) and $E_{\rm p}$ = 197$^{+173}_{-52}$~keV.}
\label{fig:gammaour}
\end{figure}

We fit the BAT total spectrum with the Band function \citep{ban}. 
Figure~\ref{fig:gammaour} shows our best fit, which gives following parameters: low-energy photon index $\alpha_{\rm B}$ = $-1.05\pm0.17$, high energy photon index
$\beta_{\rm B}= -2.3$ (fixed) and $E_{\rm p} = 197^{+173}_{-52}$~keV at $1\sigma$ and $197^{+1300}_{-70}$~keV at 90\% cl. 
We note that our first value is not consistent with the value of $E_{\rm p}=620$~keV derived by \citet{per}, and our latter value is in rough agreement with it. The discrepancy is presumably due to the neglect of soft gamma tail at $>T_0+35$~s by \citet{per}.

We compared our estimate of $E_p$ and photon index $\Gamma =1.34\pm0.08$ derived from BAT data with the empirical relation between $\Gamma$ and $E_{\rm p}$ found by \citet{zha}, their Figure~2. We find our values in excellent agreement with this relation.

To test the Amati relation, we assume the redshift of $z=1.1588$ determined by \citet{per} from the host galaxy spectroscopy.  We use standard cosmology ($H_0=70$~km~s$^{\rm -1}$~Mpc$^{\rm -1}$, $\Omega_{\rm m}=0.3$ and $\Omega_\Lambda= 0.7$), and derive $E_{\rm p,i} = 425^{+370}_{-110}$~keV  and 
$E_{\rm iso}= 7.4^{+0.1}_{-2.9} \times 10^{52}$~erg.
We find that this burst lies inside, although close to the $2\sigma$ border, of the updated Amati relation \citep[see Figure~2 in][]{ama}.

\subsection{The X-ray Afterglow}

\begin{figure}
\centering
\epsscale{1.0}
\includegraphics[angle=0, scale=0.35]{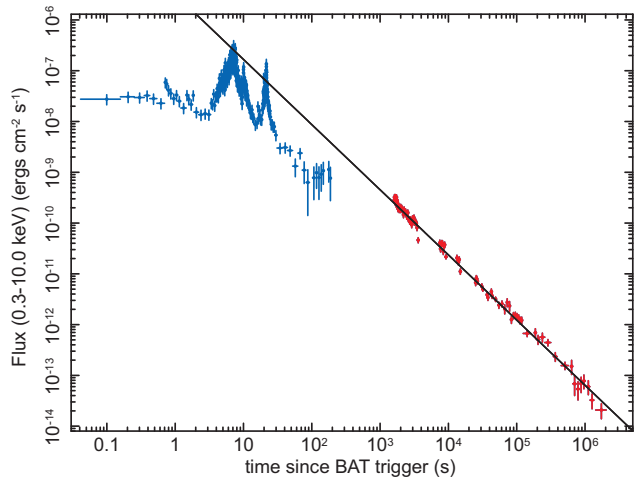}
\caption{Red: XRT light curve of GRB~061126 in $0.3-10$~keV energy band, showing a single power-law decay with small fluctuations evident throughout. The best fit power-law index is $\alpha_{\rm X}=1.290\pm0.008$. Blue: BAT light curve is shown for comparisson.
}
\label{fig:lc}
\end{figure}

\begin{figure}
\centering
\includegraphics[angle=-90, scale=0.35]{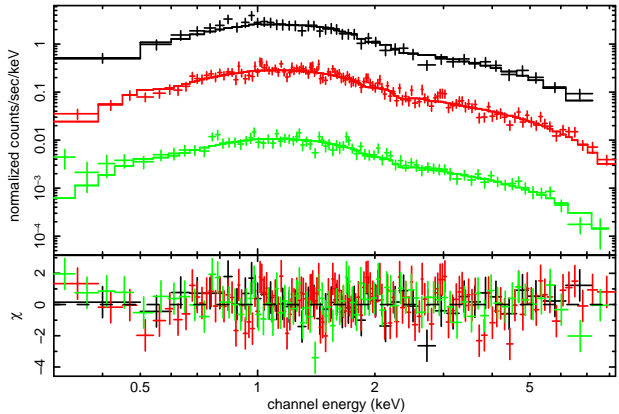}
\caption{Spectral fit of XRT data for GRB~061126. Black: data and best fit for the first sequence 
WT mode; red: data and best fit for the first sequence PC mode 
(pile-up corrected); green: data and best fit for sequences 2 to 7 PC 
mode.}\label{fig:sp}
\end{figure}

\begin{figure}
\centering
\epsscale{1.0}
\includegraphics[angle=0,scale=0.95]{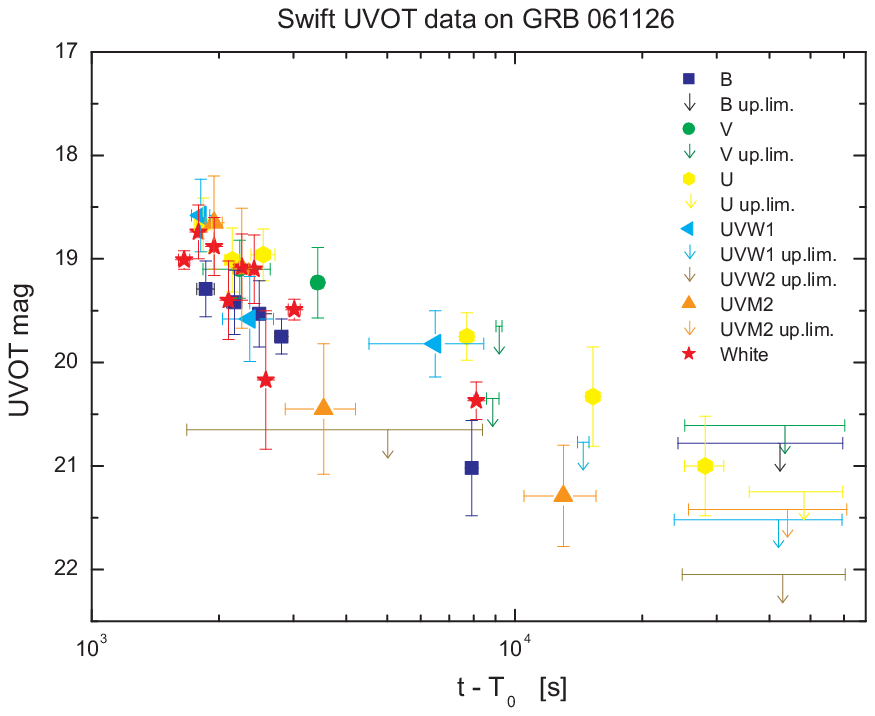}
\caption{{\em Swift} UVOT observations of GRB~061126 afterglow in $B, V, U, UVW1, UVW2, UVM2$ and White filters. Symbols show detections and arrows upper limits in given filter. Data shown in this plot are not corrected for galactic extinction.}
\label{UVOT}
\end{figure}

\begin{figure*}
\centering
\epsscale{1.0}
\includegraphics[angle=0,scale=1.35]{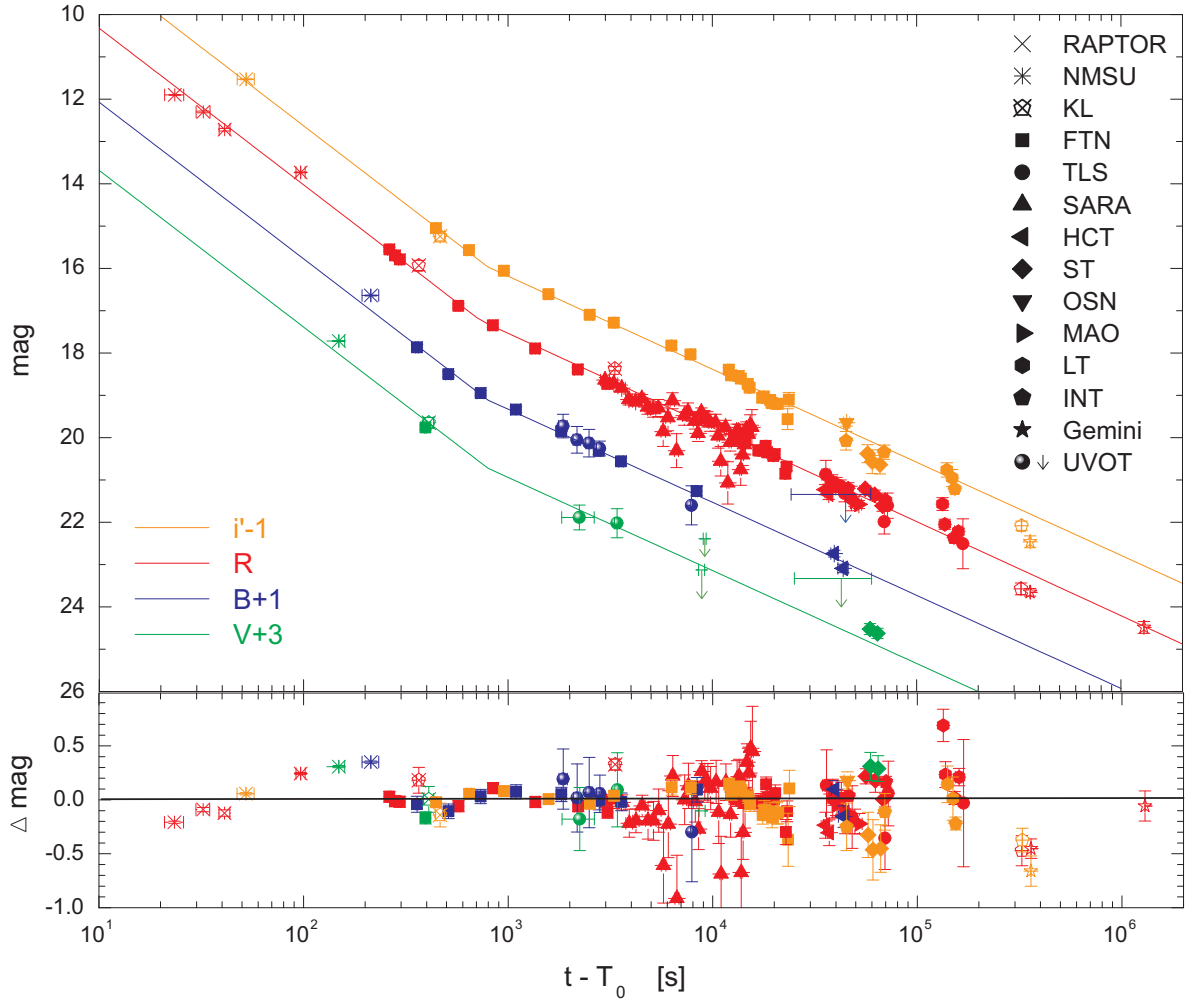}
\caption{Light curves of optical afterglow of GRB~061126 in $B, V, R$ and $i'$ bands ($r', I$ data points are reported here in $R, i'$ using \citet{smi2} filter transformations). 
Fit with a broken power law gives $\alpha_1=1.48\pm0.06$, $\alpha_2=0.88\pm 0.03$ and $t_{\rm flat}=T_0+(798\pm 53)$~s. Open symbols show data points which were excluded from the fit; 
arrows mark upper limits.}
\label{fig_lc}
\end{figure*}

\begin{figure*}
\centering
\epsscale{1.0}
\includegraphics[angle=0,scale=1.2]{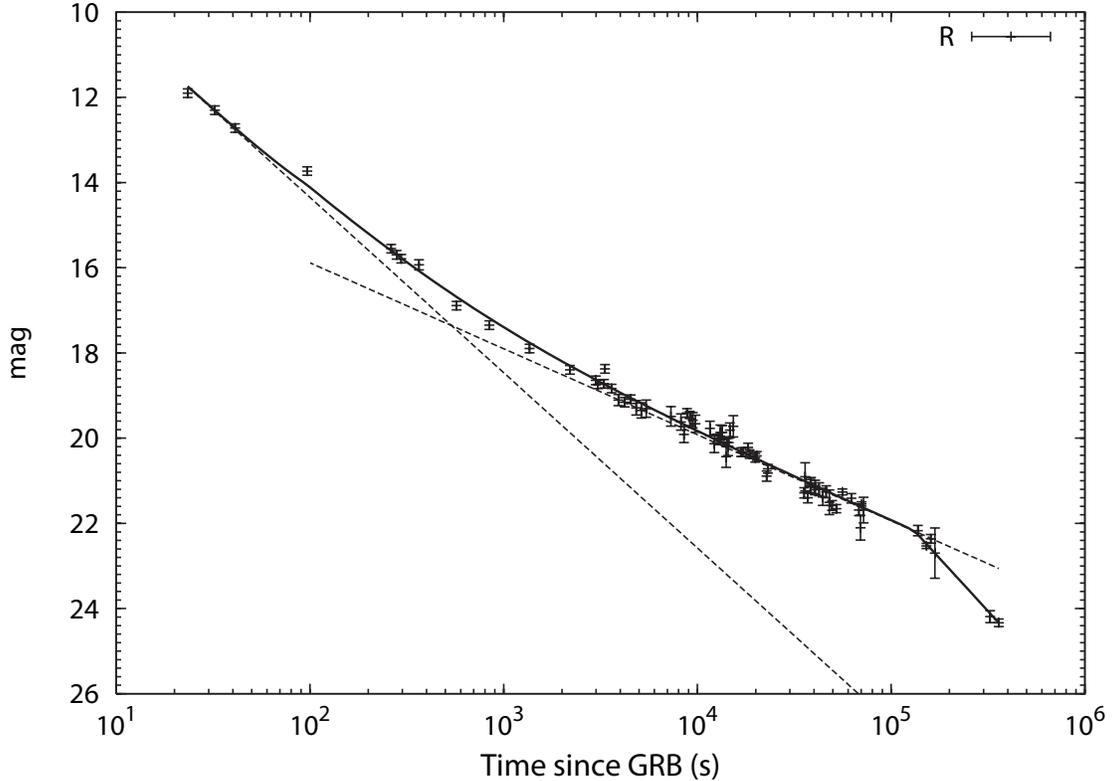}
\caption{Optical afterglow of GRB~061126 in $R$ band (corrected for host galaxy contribution, case (i), see text) and the fit with two components: reverse shock and forward shock emission giving the best fitting parameters: $\alpha_r=1.69\pm 0.09$ and $\alpha_f=0.78\pm 0.04$.
At late time we allow the forward shock component to have a break: the best fit gives $\alpha_{f,2}=1.98\pm0.15$ and time of the break $t_{\rm late-break}=(1.31\pm 0.2)\times 10^5$~s.
}
\label{fig:rsfs}
\end{figure*}

The temporal behavior of the X-ray afterglow, shown in Figure~\ref{fig:lc}, is
well described by a single power-law with index $\alpha_{\rm X}=1.290\pm0.008$; however, we note that the large value of
the $\chi^2/{\rm dof}=198.2/99$, reflects the presence of statistically significant fluctuations around the best-fitting power law. 
A similar flux variability was observed also in other X-ray afterglows, such as for GRB~060124 \citep{rom}.
In order to investigate the possibility of a hidden X-ray break \citep{cur}, 
fits with a broken or a smoothly broken power-law were performed, but they did not give a significant statistical improvement with respect to the simple power-law fit. Since the fit with a broken power law gives a slightly lower value of $\chi^2/{\rm dof}=193.7/97$, we performed F-test which showed that there is 33\% probability that this improvement is due to a chance.
Therefore, we conclude that there is no evidence for a break in the X-ray light curve up to $T_0+1\times 10^6$~s.

The three XRT spectra (from WT, first PC and later PC sequence) were fitted separately using  a power-law with a two component absorption, the first fixed at the Galactic value of $N_{\rm H}^{\rm Gal} = 1.03 \times 10^{21}$~cm$^{-2}$ \citep{kal}, the second taking into account the intrinsic absorption, left free to vary. No substantial spectral evolution was found. 
Therefore, a simultaneous fit of the 
three spectra was performed in order to improve the significance of the fit parameters 
(see Figure~\ref{fig:sp}). 
The best fit gives a power-law index $\Gamma =
1.88 \pm 0.03$, and an intrinsic absorbing column density $N_{\rm H} = 6.1 \pm 0.5 \times 10^{21}$~cm$^{-2}$.

\subsection{The UV/Optical Afterglow}\label{OptAft}

UVOT data are shown in Figure~\ref{UVOT}.  The late start of the UVOT
observations (at $T_0+26$~min) and the faintness of the afterglow at
this time resulted in large uncertainties on the measured
magnitudes. We therefore did not attempt to fit the UVOT light curves
separately. From the detection of the optical counterpart in White,
$V, B, U, UVW1$ and $UVM2$ filters, a photometric upper limit of the
redshift of GRB~061126 could be estimated to be $z\lesssim 1.5$
\citep{sba3}, which is in agreement with $z=1.158$ estimated
spectroscopically by \citet{per}.  UVOT data in $B$ and $V$ bands were
re-calibrated and combined with the ground-based observations.

The ground-based optical observations are summarized in
Table~\ref{observations}.  Light curves of the optical afterglow of
GRB~061126 in $BVRi'$ bands are plotted in Figure~\ref{fig_lc}.  They
show a power law decay with the steep-to-shallow transition between
$T_0+700$~s and $T_0+800$~s, which is apparent in all filters.  The
light curves in all four filters were fitted with the same broken
power-law, using all data points. The resulting $\chi^2/{\rm dof}$ was high
due to fluctuations in the time interval between $T_0+6\times 10^3$~s
to $T_0+2\times 10^4$~s.  
Assuming a systematic error of 0.05~mag and 0.1~mag (added in quadrature) significantly improves
the fit, giving broken power law parameters: $\alpha_1=1.48\pm0.06$,
$\alpha_2=0.88\pm0.03$ and $t_{\rm flat}=T_0+(798\pm53)$~s with
$\chi^2/{\rm dof}=278/133$ and $\chi^2/{\rm dof}=166/133$, respectively. 
The bottom panel of Figure~\ref{fig_lc} shows the
residuals with respect to the fit, where some additional variability
is still evident.

As discussed in the literature, the flattening of the optical light
curve suggests a reverse shock origin for the early steep
decay. Theoretical models \citep{kob3, zha, kob1} predict that while
the reverse shock component decays with\footnotemark[3]
$t^{-\alpha_{r}}$, the forward shock emission initially rises as
$\propto t^{0.5}$, reaches the peak at $t_p$ when the typical
frequency crosses the observation band, and decays afterwards with
$t^{-\alpha_{f}}$.  The total flux is then a sum of both components. At
earlier times, the reverse shock emission dominates the optical band,
and masks the forward shock peak. The superposition of two simple
power law components is used to fit the observational data. 
The best fitting temporal
indices are: $\alpha_r=1.69\pm 0.09$ and $\alpha_f=0.78\pm
0.04$. 
The quality of the fit, as indicated by $\chi^2/{\rm dof}=160/77$,  
does not appear to be very robust, but we believe this reflects the presence of additional fluctuations superimposed on the underlying light curve, as discussed earlier; overall, a reverse-shock, forward-shock scenario provides an adequate explanation for the underlying light curve shape.
In \S \ref{LateAft}, we develop this further by considering the effect of the late-time behavior and properly accounting for the contribution of host galaxy.

\footnotetext[3]{Throughout this paper subscripts {\it r} and {\it f} indicate reverse and forward shocks, respectively.}

\subsection{The Late-Time Afterglow and a Possible Break}
\label{LateAft}
From Figure~\ref{fig_lc} 
it is evident that there is no sign of the steepening of the optical light curve up to $T_0+1.3\times 10^5$~s. We can therefore set a firm lower limit to the time of the possible late-time break to be $t_{\rm late-break}> T_0+1.3\times 10^5$~s.

Later data points obtained by the INT and Gemini North at $\sim
T_0+3\times 10^5$~s in $r'$ and $i'$ bands (which were excluded from our earlier fits) lie
3.4 and 5.0~$\sigma $ below the best fit curves. This discrepancy
could be due to the fluctuating nature of the afterglow or indicate
the presence of late time steepening. To further investigate the
latter possibility, we considered the last optical data point (Gemini
North), at $T_0+1.3\times10^6$~s. This point seems to agree well with
our fits, however, the Gemini image shows that the OT was already faint compared to the host
galaxy. Unfortunately, it is not possible to reliably separate the
contributions of the afterglow and the host galaxy to the measured magnitude. 
Nevertheless, we can conclude that, taking into account the host galaxy contribution, the afterglow is 
fainter than what is expected in the absence of a late-time break. 
It is therefore likely that there was a steepening of the light curve before $t_{\rm
late-break}<T_0+1.3\times10^6$~s.

To further constrain the time of the possible late time break, we
considered two scenarios for the last Gemini North point: (i) it represents the
magnitude of the host galaxy only, or (ii) the magnitude of the host
galaxy and afterglow are comparable at this time. 
We corrected the afterglow
$R$-band light curve for the host contribution and repeated the above
reverse-and-forward shock fit, while allowing the forward component to
have a late time break. In both cases the best fit parameters
$\alpha_r$ and $\alpha_f$ agree with previously derived values, while
for the late time break we obtain: $t_{\rm late-break}=T_0+(1.31\pm
0.2)\times 10^5$~s in both cases, and the decay index after the break
$\alpha_{f,2}=1.98\pm0.15$ and $\alpha_{f,2}=1.38\pm0.09$ for cases
(i) and (ii) respectively.  The result for the case (i) is shown in
Figure~\ref{fig:rsfs}. Case (i) is favoured by \citet{per} measurement of the host galaxy magnitude $R=24.10\pm0.11$ at $T_0+53$~days. This is brighter than our Gemini North data point at $T_0+15$~days \citep[discrepancy is persumably due to larger aperture used by][]{per}, implying that the latter is predominantly host.

From the updated Ghirlanda correlation, i.e. eq.~(5) in \citet{nav} \footnotemark[4],
we derive $E_\gamma= 7^{+13}_{-3} \times 10^{50}$~erg
(taking into account the dispersion of the correlation), $\theta =
7.9^{+9.3\, \circ }_{-2.0}$ and $t_{\rm jet-break} = T_0 +
3.3^{+23.1}_{-2.0}$~days (assuming $n=3$, $\eta=0.2$).  This value is
consistent with our conservative estimate above: 1.5~days $< t_{\rm
late-break} - T_0<$ 15~days and with our value of $t_{\rm late-break}=
T_0+(1.52 \pm 0.23)$~days, obtained by the fits.  Nevertheless, the
interpretation of this break as due to collimation is
questionable, because there is no evidence of a simultaneous break in
the X-rays (for more examples and discussion on optical/X-ray breaks see \citet{wil} and \citet{lia}). A jet break visible only in the optical band is allowed if the X-ray emission originates
from a different emission process or an emitting region that is
physically distinct from that responsible for the optical radiation.

\footnotetext[4]{The value of 2.72 in eq. (5) in \citet{nav} should be replaced with 3.72. Private communication, L. Nava.}

\begin{figure}
\centering
\epsscale{1.0}
\includegraphics[angle=-90,scale=0.34]{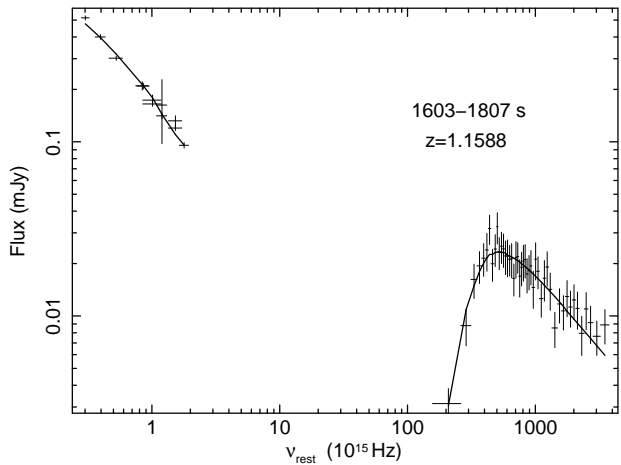}
\caption{SED after the first optical break when the light curves become shallower; at $T_0+(1600-1800)$~s. Broken power-law fit gives $\nu_{\rm br}=(4.8\pm 1.8)\times 10^{17}$~Hz,   $\beta_{\rm X}=0.88\pm 0.03$,  $\beta_{\rm O}=\beta_{\rm X}-0.5=0.38\pm 0.03$, $A_V=0.38\pm0.03$~mag, $N_{\rm H}=(8.8\pm1.2)\times 10^{21}$~cm$^{-2}$ and $\chi^2/{\rm dof}=58/53$ }
\label{SED3}
\end{figure}

\begin{figure}
\centering
\epsscale{1.0}
\includegraphics[angle=-90,scale=0.34]{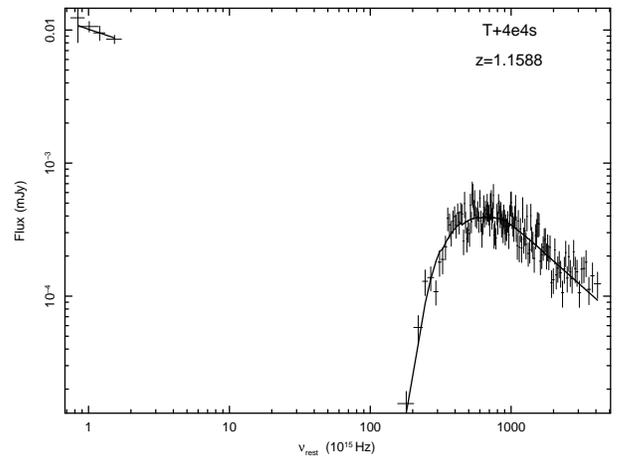}
\caption{Late time SED; at $T_0+4\times 10^4$~s. Broken power-law fit gives $\nu_{\rm br}=(9.3\pm 1.5)\times 10^{17}$~Hz, $\beta_{\rm X}=0.98\pm 0.02$,  $\beta_{\rm O}=\beta_{\rm X}-0.5=0.48\pm 0.02$, $A_V<0.13$~mag, $N_{\rm H}=(9.2\pm 0.8)\times 10^{21}$~cm$^{-2}$ and $\chi^2/{\rm dof}=152/133$ }
\label{SED4}
\end{figure}

\subsection{Spectral Energy Distributions}
\label{SED}
To further quantify the multiwavelength properties of this burst, spectral energy distributions (SEDs) at different epochs were constructed.
Optical data points at particular epochs were calculated using the interpolation with the best broken power law fit derived in \S \ref{OptAft}. 
SEDs are presented in the GRB rest frame assuming redshift $z=1.1588$. 
In all fitting procedures we applied 
the Small Magellanic Cloud (SMC) extinction profile \citep{pei} and resulting $A_V$ is the host galaxy, rest frame extinction. 

\subsubsection{Comparison of optical and X-ray emission at $T_0+50$~s and at $T_0+395$~s} \label{EarlySED}

Although the optical to X-ray SED at early time is useful for
diagnosing the magnetization of a fireball, there are no early XRT
observations. We therefore estimate the X-ray emission at early time
from a simple back-extrapolation of the monotonic decay $\alpha_{\rm
X}$ observed at later time.  This comparison is made at two different
epochs before the flattening in the optical light curve, i.e. at
$T_0+50$~s and at $T_0+395$~s. We made use of the {\em Swift} XRT data
and ground based optical data obtained by the FTN.  At both epochs we
obtain acceptable fits with a broken power law with: $\beta_{\rm
O}=\beta_{\rm X}-0.5 \approx 0.4-0.5$, $\nu_{\rm br}\approx
(2.4-5.8)\times 10^{17}$~Hz and $A_V=0.3-0.55$ mag. 

In the case of a strongly magnetized fireball, an excess of optical emission is predicted; no conclusive evidence for such an excess is detected in GRB061126.
Nevertheless, the lack of an excess is  not strongly constrained by this comparison, because even if an excess were present, the data could be equally well fitted with a different broken power law with a lower $\nu_{\rm br}$.
The use of an SED to determine the magnetization of the fireball is therefore weak in this
case and we introduce a more robust parameter to investigate magnetization in \S \ref{OptEm}.

\subsubsection{SED at $T_0+(1600-1800)$~s}

Figure~\ref{SED3} shows the SED constructed at $T_0+(1600-1800)$~s,
shortly after the optical light curves become shallower, and when the
{\em Swift} XRT observations began. We include the FTN and \citet{per}
data. A single power law does not give an acceptable
fit.  However, fitting the SED with a broken power law, assuming $\beta_{\rm
O}=\beta_{\rm X}-0.5$,  gives an acceptable fit
($\chi^2/{\rm dof}=58/53$) with the break frequency between the optical and
X-ray bands. Best fit parameters are: $\nu_{\rm br}=(4.8\pm 1.8)\times
10^{17}$~Hz, $\beta_{\rm X}=0.88\pm 0.03$, $\beta_{\rm O}=0.38\pm
0.03$, $A_V=0.38\pm0.03$~mag , and $N_{\rm H}=(8.8\pm1.2)\times
10^{21}$~cm$^{-2}$. The best fit to the SED is shown in
Figure~\ref{SED3}. Our result is not in agreement with results by \citet{per}, who find $A_V=1$~mag and $A_V \approx 0.6-0.9$~mag in their broadband fits. Nevertheless, general conclusion, that the relative optical faintness (when compared to X-ray flux) is not due to absorption, is the same.

\subsubsection{SED at $T_0+4\times 10^4$~s}
The late time SED at $T_0+4\times 10^4$~s includes the {\em Swift} XRT and ground-based optical data in $B$ (by HCT), $V$ (by ST), $R$ (by TLS, HCT, MAO and INT) and $i'$ (by S. Nevada and INT) filters. The SED is shown in Figure~\ref{SED4}, which also shows
the best fit with a broken power-law with parameters: $\nu_{\rm br}=(9.3\pm 1.5)\times 10^{17}$~Hz, $\beta_{\rm X}=0.98\pm 0.02$,  $\beta_{\rm O}=\beta_{\rm X}-0.5=0.48\pm 0.02$, $A_V<0.13$~mag, $N_{\rm H}=(9.2\pm 0.8)\times 10^{21}$~cm$^{-2}$ and $\chi^2/{\rm dof}=152/133$.

\vspace{0.5cm}
Although the uncertainty on the break frequency $\nu_{\rm br}$ derived from the fits to the SEDs at different epochs is relatively large, there is a suggestion that $\nu_{\rm br}$ is increasing with time. We discuss possible reasons for this in \S \ref{DisX}.

\subsection{Dark, gray or neither?}\label{dark}
The time of the last SED, $T_0+11$~h, is the time at which \citet{jak} compare the X-ray and optical fluxes of GRBs and define the slope of the spectral energy distribution between the optical and the X-ray band $\beta_{\rm OX}=0.5$ as dividing optically bright from optically dark bursts.
In the case of GRB~061126, at early time $\beta_{\rm OX}$ is less than this value \citep[as also noted by][]{per}: our data yield $\beta_{\rm OX}=0.29\pm0.04$ at $T_0+2000$~s, which is in slight excess of $\beta_{\rm OX}=0.23$, derived by \citet{per}. GRB~061126 could therefore be classified as a dark burst, in spite of the fact that at this early time, it is one of the optically brightest bursts detected \citep[see for example Figure~1 in][]{kan2}. However, as the afterglow is fading more slowly in the optical than in X-rays, $\beta_{\rm OX}$ is increasing with time. At $T_0+11$~h we find that $\beta_{\rm OX}=0.53\pm 0.02$, i.e. on the ``edge" of being a dark burst. 
This burst is clearly one of those for which the Jakobsson classification must be considered as a function of time and for which a simple extrapolation to/from $T_0+11$~h is inadequate \citep[for other possible cases see][]{melA}. 

Nevertheless, observations at early time give a stringent test, i.e. a value of $\beta_{\rm OX}$ below the theoretical limit of $0.5$ implies that we can not explain it in the standard fireball model.

\section{Discussion}

\subsection{Optical Emission} \label{OptEm}
The early optical behavior of GRB~061126 resembles the optical light
curves of GRB~990123 \citep{ake,nak}, GRB~021211 \citep{fox, pan} and GRB~060117 \citep{jel1}. 
We note particular similarity
with the optical afterglow of GRB~021211, not only in similar $t_{\rm flat}$,
$\alpha_1$ and $\alpha_2$, but also in cosmological redshift. As
discussed in these previous cases, the flattening behavior of light curves
can be interpreted with a reverse and forward shock scenario.  The light
curve of  GRB~061126 is composed of two segments: an
initial steep decline followed by a shallower decay with the typical
decay index of afterglow $\alpha \sim 1$. While this typical, shallower
decay is due to the forward shock (with $\alpha_f\approx 0.8$), the most
likely explanation for the early optical emission is that it is
dominated by short-lived emission from a reverse shock \citep{mes, sar2} (with $\alpha_r\approx 1.7$). The color change reported by
\citet{per} also implies the presence of different emission
components.    

The emission components of forward and reverse shocks were studied in a
unified manner by \citet{kob3} and \citet{zha}. The optical light curve of
GRB~061126 (and those of GRB~990123 and GRB~021211) is well described by
the flattening type light curve in \citet{zha}. The evolution of reverse
shocks is classified into two cases (Sari \& Piran 1995)
depending on the initial Lorentz factor of a fireball shell
$\Gamma$. The  critical value is: 
$$
\Gamma_c= [3(1+z)^3E/32\pi n m_p c^5 T^3]^{1/8},
$$  
where $E$, $T$, $n$, $z$, $m_p$ are the explosion energy,
the duration of prompt emission, the ambient matter density, the redshift
and the mass of proton, respectively.  

If $\Gamma > \Gamma_c$, the reverse shock becomes relativistic in the
frame of unshocked shell material whilst crossing the shell, and
drastically decelerates the shell (thick shell case). If $\Gamma
\lesssim \Gamma_c$, the reverse shock cannot decelerate the shell
effectively (thin shell case). Since the optical afterglow is
already fading immediately after the prompt gamma-ray emission, the
initial Lorentz factor should be comparable to or larger than the
critical value \citep{sar1}. Alternatively, if $\Gamma \gg
\Gamma_c$, the reverse shock emission should initially drop sharply
with $\alpha \sim 3$, as a rarefaction wave quickly transfers the
shell's internal energy to the ambient matter \citep{kob2,kob4}. Therefore, in the case of GRB~061126, the initial
Lorentz factor should be close to the critical value,
\begin{equation} \Gamma \sim \Gamma_c  
 = 260 ~n^{-{1\over 8}}
 \paren{\frac{1+z}{2.16}}^{3\over 8}
 \paren{\frac{T}{30 ~\mbox{sec}}}^{-{3\over 8}}
\paren{\frac{E}{7.4\times 10^{52} ~\mbox{erg}}}^{1\over 8}.
\label{eq:etac}
\end{equation}

For $\Gamma \sim \Gamma_c$, the reverse shock does not heat the shell
well. The thin shell model should be applicable to characterize the
reverse shock light curve. After the shock crossing  $t>T$, reverse
shock light curves at frequencies $\nu_{m,r} < \nu < \nu_{c,r}$ behave
as $F_\nu \propto t^{-(3p+1)/4}$, where $\nu_{m,r}$ and $\nu_{c,r}$ are
the typical and cooling frequencies of the reverse shock emission, $p$ is the electron spectral index, and a
simple approximation form is employed for the decay index \citep{zha}. 
Light curves at frequencies $\nu < \nu_{m,r}$ are shallower
as $F_\nu \propto t^{-16/35}$, and there is essentially no emission above
$\nu_{c,r}$ \citep{kob1}. The observed decay index $\alpha_r=1.69\pm0.09$ 
suggests $\nu_{m,r} < \nu < \nu_{c,r}$ during the steep decay phase and
 electron distribution index $p\sim 1.9$. As also observed in other
bursts, there is a bump feature in the optical light curve
around the break $t\sim t_{\rm flat}$ \citep{per}. A density variation in the ambient
medium is often discussed as the origin of bump features
\citep[e.g.][]{laz, guiApJ, mun07a}. If the bump is subtracted from the light curve, the
values of $\alpha$ and $p$  could be larger. 

At the shock crossing time $t\sim T$, the spectral characteristics of
the forward and reverse shock emission are related by the following
simple formulae \citep{zha},  
\begin{equation} 
\frac{\nu_{m,r}}{\nu_{m,f}} \sim \Gamma^{-2} R_B^{1/2}, \ \
 \frac{\nu_{c,r}}{\nu_{c,f}} \sim R_{B}^{-3/2},  \ \
 \frac{F_{{\rm max},r}}{F_{{\rm max},f}} \sim \Gamma R_{B}^{1/2},  \ \
 \label{eq:relations}  
\end{equation} 
where $F_{{\rm max},r}$ and $F_{{\rm max},f}$ are the peak flux of the reverse and
forward shock emission. We have assumed that $p$ and the electron equipartition parameter $\epsilon_e$ 
are the same for both the forward and reverse shock regions,
but with different magnetic equipartition parameter $\epsilon_B$ as parameterized by the magnetic energy
ratio $R_B=\epsilon_{B,r}/\epsilon_{B,f}$. Note that the definition of
the magnetization correction factor is different from that in \citet{zha}.
The reason we introduce the $R_B$ parameter is that a fireball may be
endowed with primordial magnetic fields at the central engine, so that
in principle $R_B$ could be larger than unity. As we will discuss in 
\S \ref{DisX}, the optical light curve (before and after the
break $t_{\rm flat}$) is consistent with the assumption that $p$ is the same
in the two shock regions.  

Assuming no or moderate primordial magnetization in the fireball, we
obtain a relation $\nu_{m,r} < \nu_{m,f} < \nu_{c,r} \le \nu_{c,f}$ 
at the shock crossing time. Since $\nu_{m,r}<\nu_{\rm opt}<\nu_{c,r}$
should hold during the steep decay phase, the optical band should be at 
$\nu_{m,r}<\nu_{\rm opt}<\nu_{m,f}$ or  $\nu_{m,f}<\nu_{\rm opt}<\nu_{c,r}$ at 
$t=T$. In the former case, the forward shock emission
should peak at $t=t_p$ when the typical frequency $\nu_{m,f}$ goes
through the optical band. Using $\nu_{m,f}(t_p)=\nu_{\rm opt}$ and a
scaling $\nu_{m,f}\propto t^{-3/2}$, one finds the peak time ratio as
\begin{equation}
R_t\equiv t_p/T=(\nu_{m,f}(T)/\nu_{\rm opt})^{2/3}. 
\label{eq:Rt}
\end{equation}
Following a similar discussion in
\citet{zha}, the peak flux ratio is\footnotemark[5] 
\begin{eqnarray}
R_F\equiv F_{p,r}/F_{p,f} &=& F_{{\rm max},r}(\nu_{\rm opt}/\nu_{m,r})^{-{(p-1)/2}}
/F_{{\rm max},f} \\
&=&\Gamma^{-(4\alpha_r-7)/3} R_B^{(2\alpha_r+1)/6}R_t^{\alpha_r-1},
\label{eq:Rf}
\end{eqnarray}
where $\alpha_r = (3p+1)/4$ is the decay index of reverse shock
emission, and we have used equations (\ref{eq:relations}) and (\ref{eq:Rt}).
Modifying equation (\ref{eq:Rf}), the magnetic energy ratio is given by 
\begin{equation} R_B= \left(\frac{R_F\Gamma^{(4\alpha_r-7)/3}}{R_t^{\alpha_r-1}}
    \right)^{6/(2\alpha_r+1)}.
\label{eq:Rb}
\end{equation}
In the latter case $\nu_{m,f}<\nu_{\rm opt}<\nu_{c,r}$, the forward shock
emission also peaks at $t=T$, and it follows that $R_t=1$. It is possible
to show that equation (\ref{eq:Rb}) is still valid. 

\footnotetext[5]{This ratio $R_F$ differs from ${F_{{\rm max}, r}}/{F_{{\rm max}, f}}$ defined in eq.~\ref{eq:relations}: $F_{\rm max}$ is a peak flux in the spectral domain at a given time, while $F_p$ is a peak flux in the time domain at a given frequency.}

This event is a marginal case with $\Gamma \sim \Gamma_c$, and the
reverse shock emission should peak around the end of the prompt
gamma-ray emission at $t \sim 30$~s. Unfortunately the forward shock
peak $t_p$ was not caught (it means that $t_p \lesssim  t_{\rm flat} \sim
800$~s), because the reverse shock emission dominated at early times. First 
we consider the case with the upper limit $t_p =t_{\rm flat}$, and estimate
$R_B$. The peak time ratio is $R_t \sim 27$. 
Since the optical light curve flattens at $t_{\rm flat} (\sim t_p)$, the
reverse shock and forward shock components are comparable at that time:
$F_r(t_p)= F_{p,r} (t_p/T)^{-\alpha_r} \sim F_{p,f}$, 
and it follows that the peak flux ratio can be written as $R_F \sim
R_t^{\alpha_r}$. Substituting this relation into equation (\ref{eq:Rb}) 
we obtain 
\begin{equation}
R_B \sim \paren{R_t^3
\Gamma^{(4\alpha_r-7)}}^{2/(2\alpha_r+1)} \sim 50 
 \ \ \mbox{for} \ \  \alpha_r=1.69,
\end{equation}
where ambient matter density $n=1$~proton/cm$^3$ is assumed, but the 
result is insensitive to $n$ as $R_B \propto n^{0.01}$ for $\alpha_r=1.69$. 
If the forward shock emission reaches the maximum earlier
$t_p<t_{\rm flat}$, the value of $R_B$ might be different. To evaluate how
$R_B$ depends on $t_p$, we refer to scalings: $R_t\propto t_p$ and 
$R_F\propto F_{p,f}^{-1} \propto t_p^{\alpha_f}$, where we took into
account that the peak of the forward shock emission should be on the
power law line with $\alpha_f$. Using these scalings, one finds that the
dependence is weak\footnotemark[6]:
$R_B \propto t_p^{6(1-\alpha_r+\alpha_f)/(1+2\alpha_r)} \propto t_p^{0.12}$
for $(\alpha_r, \alpha_f)=(1.69,0.78)$. In the earliest case  $t_p\sim
30$ s, we obtain $R_B \sim 34$. 
These results imply that magnetic energy density in a fireball is much 
larger than in the forward shock. Nevertheless, as a small value of $\epsilon_{B,f}\sim 10^{-4}-10^{-2}$ is
usually inferred from afterglow modeling \citep[e.g.][]{pankum}, the above values of $R_B$ suggest that $\epsilon_{B,r}\ll 1$ and the energy in the fireball is still likely to take the form 
of kinetic energy (a baryonic fireball) rather than Poynting flux.

\footnotetext[6]{If the decay indices of the forward-shock and reverse-shock 
emission exactly satisfy the theoretical values: $\alpha_f=3(p-1)/4$ and
$\alpha_r=(3p+1)/4$, a relation $\alpha_r-\alpha_f=1$ should hold,
and $R_B$ does not depend on $t_p$.}

Recently the Liverpool Telescope obtained an early-time estimate of
optical polarization of a GRB afterglow shortly after the burst
\citep{mun07b}. Polarization observations of an afterglow with a
flattening light curve or at the time of a reverse shock peak would
provide additional constraints on the presence of magnetized fireballs.
 
\subsection{X-ray Emission}
\label{DisX}
The X-ray afterglow was observed from $\sim T_0+26$~min to $T_0+20$~days by {\em Swift} XRT. 
The X-ray light curve fades as a single
power law with a decay index $\alpha_{\rm X} = 1.290\pm0.008$. Since this is
steeper than the optical light curve $\alpha_f\sim 0.8$ over the
same period, the X-ray band should be in a different spectral domain
than the optical, therefore: $\nu_{m,f} < \nu_{\rm opt} < \nu_{c,f} < \nu_{\rm X}$.  
The observed X-ray spectral index $\beta_{\rm X}=0.94\pm0.05$ corresponds to
$p=2 \beta_{\rm X} \sim 2.0$. The optical emission from the reverse shock and
forward shock should decay as $\alpha_r=(3p+1)/4=1.75$ and 
$\alpha_f=3(p-1)/4=0.75$, respectively. These are in good agreement with the observed $\alpha_r=1.69\pm0.09$ and
$\alpha_f=0.78\pm 0.04$. As we have discussed in the previous section,
the bump feature in the early optical light curve might make the
observed $\alpha_r$ smaller and the observed $\alpha_f$ larger.  

The observed emission in X-rays ($\alpha_{\rm X}= 1.29$) decays
faster than that expected from the X-ray spectral
index $\alpha_{\rm X}=(3p-2)/4=1.0$. This discrepancy might be due to the
radiative loss. If the energy distribution of electrons is flat $p\sim
2$, each decade in the electron distribution contains the same amount of
energy, and if $p<2$, high energy electrons have most of the total
electron energy. Even in the slow cooling regime $\nu_{m} < \nu_{c}$,
the radiative loss might make the decay steeper $\Delta \alpha \sim
\epsilon_e$ \citep{sar1}. However, the radiation loss affects the optical emission also, and 
       the expected (steeper) optical index is not consistent with the
       observations. More general discussion on the time dependent
       parameters can be given as follows.
       The ratio of X-ray to optical flux depends 
       on parameters as 
      \begin{equation}
      F_{\rm X}/F_{\rm opt} \propto \nu_{c,f}^{1/2} \propto \epsilon_{B,f}^{-3/4}
      E^{-1/4} n^{-1/2} t^{-1/4}
      \end{equation}
      where we have assumed the standard synchrotron spectrum
      ($\nu_{m,f}<\nu_{\rm opt}<\nu_{c,f}<\nu_{\rm X}$) and an adiabatic evolution
      of the blast wave. If the parameters are slowly changing in time 
      (e.g. radiation loss, late-time energy injection, 
        time dependent microscopic parameters or a gradient in the ambient 
        density), the difference of the decay indexes in the
      two bands could be larger than  the standard value
      $\Delta \alpha=\alpha_{\rm X}- \alpha_{\rm O}=1/4$.  The observed difference $\Delta \alpha=0.51$ 
      requires that the parameters should increase with time. This is
      a somewhat unphysical condition, and causes a large discrepancy 
      between the theoretical and observed optical decay indices, because
      the optical flux is sensitive to the parameters as
      $F_{\rm opt}\propto \epsilon_e^{p-1} \epsilon_{B,f}^{(p+1)/4}
      E^{(p+3)/4}n^{1/2}t^{-3(p-1)/4}$.
      Inverse Compton scattering can, in principle, affect the 
          cooling frequency $\nu_{c,f}$, and a correction factor to 
          $\nu_{c,f}$ is time-dependent during the slow cooling 
          phase. However, the presence of strong Inverse Compton cooling 
    will make the difference $\Delta \alpha$ even smaller \citep{sar3}.

An alternative possibility for the production of the X-ray afterglow is a continued activity of the central engine 
\citep[e.g. late prompt emission,][]{ghi, fan} or a two-component jet (\citet{rac} and the references therein). The optical light curve shows evidence of a late time break which 
is consistent with the value derived from the Ghirlanda relation. 
The lack of a simultaneous (jet) break in the X-ray light curve 
might indicate that the X-rays originate from a different emission site.
Another indication is the low value of $\beta_{\rm OX}$ at early times (see \S \ref{dark}), which might suggest the enhancement of the 
X-ray flux due to the additional emission \citep{per, melA}.
If the additional emission (e.g. late-time prompt emission or narrow jet emission in the two-component jet model) masks the forward shock emission in X-ray 
band, the decay rate is determined by the process related to the 
additional component. In principle the decay rate could be faster 
than that implied by the fireball model until the forward shock 
emission is eventually unmasked. To interpret the chromatic afterglow of GRB~080319B, \citet{rac} suggest a two-component jet model in which the additional component (emission from a spreading narrow jet) decays faster than the underlying component (wide jet emission) responsible for the optical component. In the case of GRB~061126, the faster decay in X-ray could be explained if the electron distribution index $p$ is not universal and larger in the narrow jet, though we need to explain what causes the difference of $p$ in the two jets. Furthermore, the X-ray (narrow-jet emission) light curve does not show a jet break before $T_0+10^6$~s, while the optical (wide-jet emission) light curve shows a possible steepening at $t< T_0+3\times10^5$~s.
The two-component jet model might be disfavored to explain GRB~061126. As \citet{per} have discussed, any additional component models might share a difficulty to explain how to avoid contaminating the blue end of the observed optical spectrum with emission from the low-energy tail of the additional emission (a synchrotron-like spectrum) peaking at X-ray wavelengths. This could require fine-tuning or tight constraints on the additional emission model (e.g. the self absorption frequency of the additional emission is higher than the optical band).

As we noted in \S \ref{SED}, our fits suggest that $\nu_{\rm br}$ is increasing with time. This could imply the presence of a wind environment.
We have assumed a homogeneous ambient medium (ISM) in the above
discussion. Even in a wind medium, the relations (\ref{eq:relations}) are
valid. The same relation $\nu_{m,r} < \nu_{\rm opt} < \nu_{c,r}$  should be
satisfied to explain the steep optical decay in the context of the
reverse shock emission. At late times $t>t_{\rm flat}$, if the optical and
X-ray afterglows are due to the forward shock, the X-ray decay rate
should be the same as or slower than the optical decay rate,
because the cooling frequency $\nu_{c,f}$ moves blue-ward in the wind
model. This is inconsistent with the observations. Therefore, the ISM
model is favored.

One possible explanation for the increase of the break frequency would be related to the reverse shock emission component at early time. 
When the relation $\nu_{opt} < \nu_{c,f} < \nu_{X}$ is satisfied, the forward shock emission spectrum (optical to X-ray) is fitted with a broken power law. Even if reverse shock emission dominates the optical band, the spectrum could be still well fitted with a broken power law with a lower break frequency, because we observe only optical and X-ray fluxes and our samplings are spares in the frequency domain.
The reverse shock component decays faster than the forward shock component, and it becomes less prominent at late times. The break frequency should increase and approaches the break frequency of the forward shock at late times. This might explain the observed behavior. However, at the time the X-ray observations started and onwards, the contribution of the reverse shock to the optical emission is already negligible and could not significantly influence the break frequency. 
Another possibility is that of late-time prompt emission
i.e. X-rays coming from an emission site other than that of the forward shock.

The gamma-ray tail is described by a power law with $\alpha_\gamma=1.3$.
The similarity of $\alpha_\gamma$ with $\alpha_{\rm X}$ index leads to a
speculation that the $\gamma$ tail is due to the same forward shock and
that the gamma-ray band at the early times $t> T_0+30$~s and the X-ray
band at late times $t> T_0+26$~min are in the same spectral domain:
$\nu_{c,f} < \nu_{\rm X}, \nu_\gamma$. However, using $\alpha=1.3$, the
extrapolated value in gamma-ray band at $t=T_0+26$~min gives a spectral
index $\beta_{{\rm X}-\gamma} \sim 0.2$, which is much shallower than the
observed x-ray spectrum $\beta_{\rm X}\sim 1$. We therefore suspect that the
same decay index $\alpha_\gamma\sim \alpha_{\rm X}\sim 1.3$ happened by
chance. The bright gamma-ray tail might be produced by the superposition
of internal shock emission (central engine activity) or the propagation of
the forward shock in a higher density ambient medium (if so, the cooling
frequency should be above the gamma-ray band at $t<T_0+200$~s).

\section{Conclusions}
\label{sec:conc}

GRB~061126 was a long burst with intriguing optical and X-ray afterglows.
The optical light curve shows a steep-to-flat transition at about 13~min after the trigger. We showed that the early, steep component can be interpreted as due to the reverse shock ($\alpha_r=1.69\pm0.09$), while the later slowly fading component as coming from the forward shock ($\alpha_f=0.78\pm0.04$). 
From the afterglow properties we deduce that $\Gamma \sim \Gamma_c \sim 260$ and estimate the magnetic energy ratio to be $R_B \sim 34-50$. 
This indicates that the magnetic energy density in the fireball 
 is much larger than in the forward shock at the fireball deceleration, 
 but that the fireball is still likely to be baryonic and not Poynting 
 flux dominated.

The standard fireball model can explain the optical decay indices before and after the flattening, i.e.
$\alpha_1$ and $\alpha_2$, and the X-ray spectral index $\beta_{\rm X}$ with a single value of electron
index $p\sim2$. 
However,
the X-ray decay index $\alpha_{\rm X}=1.290\pm0.008$ deviates from the expected value
$\alpha_{\rm X}=1.0$. We investigated the generalized standard fireball model with time dependent parameters (e.g. radiation loss,
late time energy injection, time dependent microscopic parameters or a gradient in the ambient density),
and we found that none of these modified models can explain the observed decay and spectral indices
in a consistent manner. This could imply the presence of late-time prompt emission and a different origin of the X-ray afterglow, which would also be a possible explanation for the large ratio of X-ray to optical fluxes (i.e. shallow spectral index from optical to X-ray band) and for the possible chromatic jet break at $T_0+1.5$~days.
Although there are  significant fluctuations 
in the observed X-ray light curve, the late time internal-shock model 
could require a fine tuning of the central engine to explain the power-law decay.

\section{Acknowledgements}

AG thanks Slovenian Research Agency and Slovenian Ministry for Higher Education, Science, and Technology for financial support.
C.G., V.M. and B.S. acknowledge support from ASI grant I/011/07/0. 
CGM acknowledges financial support from the Royal Society and Research
Councils U. K. 
{\it RoboNet-1.0} was supported by PPARC and STFC. {\em Swift} mission is funded in the UK by STFC, in Italy by ASI, and in the USA by NASA.  
The Faulkes Telescopes are operated by the Las Cumbres Observatory.
The Liverpool Telescope is owned and operated by Liverpool John Moores University.
We thank the anonymous referee for useful comments and suggestions.

\clearpage

\small{

\setcounter{table}{0}
\begin{table}
 \caption{Summary of Optical Observations of GRB~061126 with the Faulkes Telescope North (FTN), Tautenburg  Schmidt telescope (TLS), 
 SARA, Sampurnanand Telescope (ST), Himalayan Chandra Telescope (HCT), 1.5-m telescope at Observatorio de Sierra Nevada (OSN),
 1.5-m telescope at Maidanak Observatory (MAO), Liverpool Telescope (LT), Isaac Newton Telescope (INT) and Gemini North.
 $T_{\rm start}$, $T_{\rm end}$ and $\Delta T_{\rm mean}$ are reffered to trigger time $T_0$. $\Delta T_{\rm mean}$ is defined as ${\sum_i (t_i \Delta t_i)}\over {\sum_i (\Delta t_i)}$, where $t_i$ is the mid time of individual exposures and $\Delta t_i$ is the exposure length.}
 \label{observations}
 \begin{tabular}{llccccc}
  \hline
Telescope & Filter & $\Delta T_{\rm mean}$ &  Mag $\pm$ Err & $T_{\rm start}$ & $T_{\rm end}$  & $T_{\rm exp}$ \\
          &        &  (min)            &                & (min)       & (min)      & (s) \\
\hline 
FTN	& $R_{\rm C}$  &   4.38 &  15.55 $\pm$ 0.05  &     4.30  &    4.47         &   10     \\
& $R_{\rm C}$ &   4.67 &  	 15.70  $\pm$ 0.05   &    4.59   &   4.76	  &    10  \\
& $R_{\rm C}$ &   4.93 &  	 15.79  $\pm$ 0.05   &    4.85   &   5.02	  &    10  \\
& $R_{\rm C}$ &   9.53 &  	 16.89  $\pm$ 0.05   &   9.28    &   9.78          &   30     \\
& $R_{\rm C}$ &  14.1  & 		 17.35  $\pm$ 0.05   & 	13.6   & 	 14.6    & 	 60  \\
& $R_{\rm C}$ &  22.7  & 		 17.89  $\pm$ 0.05   & 	21.7   & 	 23.7    & 	 120 \\  
& $R_{\rm C}$ &  36.7  & 		 18.39  $\pm$ 0.05   & 	35.2   & 	 38.2    & 	 180 \\  
& $R_{\rm C}$ &  51.1  & 		 18.73  $\pm$ 0.06   & 	50.1   & 	 52.1    & 	 120 \\ 
& $R_{\rm C}$ &212.7   & 		 19.99  $\pm$ 0.06   & 	210.1  & 	 215.3  & 	 300 \\
& $R_{\rm C}$ &218.0   & 		 20.03  $\pm$ 0.06   & 	215.4   & 	 220.6  & 	 300 \\
& $R_{\rm C}$ &236.5   & 		 20.08  $\pm$ 0.06   & 	233.9  & 	 239.2  & 	 300 \\
& $R_{\rm C}$ &241.9   & 		 20.08 $\pm$ 0.06   & 	239.3  & 	 244.5  & 	 300 \\
& $R_{\rm C}$ &280.3   & 		 20.31  $\pm$ 0.07   & 	277.8  & 	 283.0  & 	 300 \\
& $R_{\rm C}$ &285.7   & 		 20.31  $\pm$ 0.07   & 	283.1  & 	 288.3  & 	 300 \\
& $R_{\rm C}$ &304.3   & 		 20.20  $\pm$ 0.07   & 	301.7  & 	 306.9  & 	 300 \\
& $R_{\rm C}$ &309.6   & 		 20.35  $\pm$ 0.08   & 	307.0  & 	 312.2  & 	 300 \\
& $R_{\rm C}$ &331.7   & 		 20.44  $\pm$ 0.08   & 	329.1  & 	 334.3  & 	 300 \\
& $R_{\rm C}$ &337.1   & 		 20.39  $\pm$ 0.08   & 	334.4  & 	 339.7  & 	 300 \\
& $R_{\rm C}$ &379.6   & 		 20.85  $\pm$ 0.12   & 	377.0  & 	 382.2  & 	 300 \\
& $R_{\rm C}$ &384.9   & 		 20.69  $\pm$ 0.10   & 	382.3  & 	 387.5  & 	 300 \\
FTN &  $i'$ &    7.41  & 16.05 $\pm$ 0.04  &  7.33  &   7.50  &  10   \\
    &  $i'$ &	  10.75  & 16.57 $\pm$0.04   & 10.50  &  11.00  &  30	 \\
    &  $i'$ &   15.93  & 17.06 $\pm$ 0.04  & 15.43  &  16.43  &  60	 \\
    &  $i'$ &	  26.28  & 17.61 $\pm$0.04  & 25.28  &  27.28  & 120	 \\
    &  $i'$ &   41.88  & 18.10 $\pm$ 0.04  & 40.38  &  43.38  & 180	 \\
    &  $i'$ &	  54.95  & 18.28 $\pm$0.05  & 53.95  &  55.95  & 120	 \\
    &  $i'$ &  105.2  & 18.82 $\pm$ 0.07  & 104.7 &  105.7 &   60	 \\
    &  $i'$ &	 130.3  & 19.03 $\pm$0.06  & 129.3 &  131.3 &  120  \\
    &  $i'$ &  200.7 &  19.39$\pm$  0.05 &  198.1&  203.3&  300	 \\
    &  $i'$ &	 206.1 & 19.53 $\pm$0.06   & 203.5 & 208.7 & 300	 \\
    &  $i'$ &  224.6 &  19.55$\pm$  0.06 & 222.0 & 227.2 & 300	 \\
    &  $i'$ &	 230.0  & 19.61 $\pm$0.06   & 227.3 & 232.6 & 300     \\
    &  $i'$ &  248.5 &  19.73$\pm$  0.06 & 245.9 & 251.1 & 300	 \\
    &  $i'$ &	 253.8 & 19.82 $\pm$0.07   & 251.2 & 256.4 & 300	 \\
    &  $i'$ &  292.3 &  20.06$\pm$  0.08 & 289.7 & 294.9 & 300	 \\
    &  $i'$ &	 297.7 & 20.03 $\pm$0.08   & 295.1 & 300.3 & 300	 \\
    &  $i'$ &  319.8 &  20.11$\pm$  0.09 & 317.2 & 322.4 & 300	 \\
    &  $i'$ &	 325.1 & 20.19 $\pm$0.09   & 322.5 & 327.7 & 300	 \\
    &  $i'$ &	 343.7 & 20.20 $\pm$0.10   & 341.1 & 346.3 & 300	 \\
    &  $i'$ &	 349.0 & 20.20 $\pm$0.10    &346.4  &351.6  &300	 \\
    &  $i'$ &	 389.8 & 20.56 $\pm$0.25    &389.0  &390.6  &100	 \\
    &  $i'$ &	 395.1 & 20.10 $\pm$0.17    &394.3  &395.9  &100	 \\
FTN & $B$ &     5.98  & 16.87 $\pm$0.07 &   5.90 & 6.07    &  10    \\   
    & $B$ &     8.52  & 17.50 $\pm$0.07 &  8.27  & 8.77    &  30	\\
    & $B$ &    12.25  & 17.95  $\pm$0.07 &  11.75 & 12.75   &  60	\\
    & $B$ &    18.22  & 18.34 $\pm$0.06 &  17.22 & 19.22   & 120	\\
    & $B$ &    30.35  & 18.85 $\pm$0.07 &  28.85 & 31.85   & 180	\\
    & $B$ &    46.42  & 19.32 $\pm$0.08 &  45.42 & 47.42   & 120	\\
    & $B$ &    59.55  & 19.56 $\pm$0.08 &  58.05 & 61.05   & 180	\\
    & $B$ &   139.8  & 20.27 $\pm$0.11  &  138.3&  139.8 &  180	\\
FTN & $V$ & 6.58  & 16.76 $\pm$ 0.06 &  6.50 & 6.67  &    10 \\    
\hline 
TLS 	& $R_{\rm C}$ & 599.4 & 	 20.87  $\pm$  0.33 & 	  572.7  & 	 626.0 & 	 2400 \\
	& $R_{\rm C}$ &642.8 & 	 21.06  $\pm$ 0.18  & 	 626.8   & 	658.7   & 	1800  \\
	& $R_{\rm C}$ &675.4 & 	 21.11  $\pm$ 0.17  & 	 659.6  & 	691.2  & 	1800  \\
	& $R_{\rm C}$ &707.8  & 	 21.15  $\pm$ 0.15  & 	 692.0  & 	723.7  & 	1800  \\
	& $R_{\rm C}$ &740.3 & 	 21.32  $\pm$ 0.20  & 	 724.5  & 	756.1  & 	1800  \\
	& $R_{\rm C}$ &799.8  & 	 21.44  $\pm$ 0.29  & 	 778.6  & 	821.0  & 	2400  \\
	& $R_{\rm C}$ &1159 & 	 21.99  $\pm$ 0.29  & 	 1143  & 	1175   & 	1800  \\
	& $R_{\rm C}$ &1202 & 	 21.61  $\pm$ 0.30  & 	 1175  & 	1229   & 	3000  \\
	& $R_{\rm C}$ &2806 & 	 22.51  $\pm$ 0.59  & 	 2776  & 	2836  & 	3600  \\
\hline 	
SARA & $R_{\rm C}$ &      49.80  &18.64 $\pm$0.06 &47.30     & 52.30   &300 	\\
     & $R_{\rm C}$ &      54.90  &18.72 $\pm$0.08 & 52.40    & 57.40   &300	\\									      
     & $R_{\rm C}$ &	60.05  &18.83 $\pm$0.07 & 57.55    & 62.55   &300	\\
     & $R_{\rm C}$ &	65.13  &19.10 $\pm$0.13 & 62.63    & 67.63   &300	\\
     & $R_{\rm C}$ &	70.25  &19.15 $\pm$0.11 & 67.75    & 72.75   &300	\\
  \hline 
   \end{tabular}							
  \end{table}							
  \begin{table}
   \begin{tabular}{llccccc}
    \hline
  Telescope & Filter & $\Delta T_{mean}$ &  Mag $\pm$ Err & $T_{start}$ & $T_{end}$  & $T_{exp}$ \\
            &        &  (min)            &                & (min)       & (min)      & (s) \\
  \hline 

 SARA     & $R_{\rm C}$ &	75.38  &19.08 $\pm$0.10 & 72.88    & 77.88   &300	\\
     & $R_{\rm C}$ &	80.51  &19.28 $\pm$0.17 & 78.01    & 83.01   &300	\\
     & $R_{\rm C}$ &	85.62  &19.34 $\pm$0.18 & 83.12    & 88.12   &300	\\
	& $R_{\rm C}$ &	90.77  &19.30 $\pm$0.20 & 88.27    & 93.27   &300	\\
     & $R_{\rm C}$ &	95.85  &19.86 $\pm$0.35 &  93.35   & 98.35   &300  \\
     & $R_{\rm C}$ &	101.2 &19.53  $\pm$0.31 &  98.65   & 103.7  &300  \\
     & $R_{\rm C}$ &	106.3 &19.13 $\pm$0.19 &  103.8  & 108.8  &300  \\
     & $R_{\rm C}$ &	111.4 &20.31 $\pm$0.40 &  108.9  & 113.9  &300  \\
     & $R_{\rm C}$ &	121.6 &19.48 $\pm$0.23 &  119.1  & 124.1  &300  \\
     & $R_{\rm C}$ &	126.8 &19.39 $\pm$0.21 &   124.3 & 129.3  &300  \\
     & $R_{\rm C}$ &	137.0 &19.61 $\pm$0.19 & 134.5   & 139.5  &300 \\
     & $R_{\rm C}$ &	142.1 &19.90 $\pm$0.19 &  139.6  &  144.6 & 300 \\
     & $R_{\rm C}$ &	147.2 &19.40 $\pm$0.10 & 144.7   &  149.7 & 300 \\
     & $R_{\rm C}$ &	152.5 &19.53 $\pm$0.17 & 150.0   &  155.0 & 300    \\
 & $R_{\rm C}$ &	157.6 &19.57 $\pm$0.18 & 155.1   &  160.1 & 300	  \\
     & $R_{\rm C}$ &	162.8 &19.65 $\pm$0.20 & 160.3   &  165.3 & 300	  \\
     & $R_{\rm C}$ & 	173.0 &19.65 $\pm$0.20 & 170.5   &  175.5 & 300 \\ 
     & $R_{\rm C}$ &      178.1 &19.96 $\pm$0.29 & 175.6   &  180.6 & 300    \\    
     & $R_{\rm C}$ &	183.3 &20.56 $\pm$0.35 &    180.8&   185.8&  300   \\
     & $R_{\rm C}$ &	193.5 &19.76 $\pm$0.17 &    191.0&   196.0&  300   \\
     & $R_{\rm C}$ &	198.6 &21.07 $\pm$0.50 &   196.1 &  201.1 & 300    \\
     & $R_{\rm C}$ &	203.9 &20.11 $\pm$0.21 &   201.4 &  206.4 & 300 	  \\
     & $R_{\rm C}$ &	219.2 &19.85 $\pm$0.17 &    216.7&   221.7&  300   \\
     & $R_{\rm C}$ &	224.4 &19.85 $\pm$0.16 &   221.9 &  226.9 & 300  \\
     & $R_{\rm C}$ &	229.5 &20.76 $\pm$0.37 &   227.0 &  232.0 & 300  \\
     & $R_{\rm C}$ &	234.6 &20.41 $\pm$0.25 &   232.1 &  237.1 & 300 	\\
     & $R_{\rm C}$ &	239.7 &20.17 $\pm$0.22 &    237.2&   242.2&  300 \\
     & $R_{\rm C}$ &	244.9 &19.80 $\pm$0.17 &   242.4 &  247.4 & 300 	\\
     & $R_{\rm C}$ &	250.0 &19.92 $\pm$0.24 &   247.5 &  252.5 & 300  \\
     & $R_{\rm C}$ &	255.3 &19.71 $\pm$0.25 &  252.8  &  257.8 & 300 	\\
     & $R_{\rm C}$ &	260.4 &19.76 $\pm$0.42 &    257.9&   262.9&  300 \\
\hline 
ST   & $R_{\rm C}$ &  934.4 & 21.21  $\pm$ 0.07 & 919.0  & 946.1  & 1200 \\
     & $R_{\rm C}$ &   1039 & 21.35  $\pm$ 0.11 & 1024 & 1054 & 1800 \\								
    & $R_{\rm C}$ &   1136 & 21.61   $\pm$ 0.13 & 1121 & 1151 & 1800 \\
ST   & $I$ &   958.5  &21.02  $\pm$ 0.21  &948.5    &968.5   &1200  \\
     & $I$ &   1012 & 21.21 $\pm$ 0.28 & 1002  &1022  &1200  \\
     & $I$ &    1103&  21.28$\pm$  0.22&  1088 & 1118 & 1800 \\
ST   & $V$ &   985.2  &21.53  $\pm$0.13  &970.2  &1000  &1800 \\
     & $V$ &   1071 & 21.63 $\pm$ 0.12 & 1056&  1086&  1800  \\ 
\hline 
HCT   & $R_{\rm C}$ & 593.4 & 21.23 $\pm$ 0.12  & 578.3 & 604.6 & 1080 \\
      & $R_{\rm C}$ & 619.1 & 21.34 $\pm$ 0.12  & 607.8 & 630.5 & 1200 \\
HCT   & $B$ &  658.3  &21.74 $\pm$ 0.08  & 633.5  &682.9  & 2700  \\  
      & $B$ &  728.3 & 22.09 $\pm$  0.10 & 685.9 & 770.8 & 4500  \\  
\hline 
MAO  & $R_{\rm C}$ &     638.5  & 21.01 $\pm$ 0.09 &  623.5 & 653.9 & 1200 \\
     & $R_{\rm C}$ &     668.8  & 21.16 $\pm$ 0.10 &  654.5 & 682.9 & 1500 \\
     & $R_{\rm C}$ &     825.7  & 21.51 $\pm$ 0.11 &  811.5 & 840.9 & 1500 \\
     & $R_{\rm C}$ &     868.8  & 21.58 $\pm$ 0.10 &  842.0 & 895.8 & 2700 \\
\hline 
OSN  & $I$ & 757.7 & 20.68  $\pm$ 0.08 & 730.0 &  783.3 &  2700 \\
\hline 
LT 	& $r'$ & 2233 & 21.69 $\pm$ 0.15    &  2218   &  2247 &  1500 \\
	& $r'$ & 2289 & 22.17 $\pm$ 0.12	&2248 & 2329  &  4800  \\
   	& $r'$ & 2669 & 22.34 $\pm$  0.08 &	2539     & 2763   &  10800   \\
LT 	& $i'$ &  2343 & 21.76 $\pm$  0.17 & 2333 & 2354 & 1200  	 \\
	& $i'$ &  2490 & 21.95 $\pm$  0.20 & 2459  & 2520 & 3600   	 \\
\hline 
INT & $r'$   	&  770.4  & 21.33 $\pm$ 0.14 & 762.3   &778.6  &900 \\
    & $r'$	&  1171 & 21.60 $\pm$ 0.05 & 1166  &1176 & 900 \\
    & $r'$	&  2537 & 22.49 $\pm$ 0.06 &  2525 & 2548 &  1200 \\
    & $r'$	&  5404 & 23.70 $\pm$ 0.14 & 5390  &5418 & 1200 \\
INT & $i'$   	&  753  & 21.07 $\pm$  0.22 &  744.6 & 760.9  & 900  \\
    & $i'$	&  1156 & 21.34 $\pm$  0.17 & 1148 & 1164 & 900  \\
    & $i'$	&  2565 & 22.21 $\pm$  0.06 & 2549 & 2580 & 1500 \\ 
    & $i'$	&  5431 & 23.08 $\pm$  0.11 & 5420 & 5443 & 1200 \\ 
INT & $g'$   & 2593 & 22.96 $\pm$ 0.06 & 2581 & 2604 & 1200  \\
    & $g'$   & 5457 & 23.91 $\pm$ 0.11 & 5445 & 5468&  1200 \\
\hline 
Gemini North & $r'$  & 5993  & 23.78 $\pm$ 0.09 & 5990  & 5996  & 360	 \\
       & $r'$	&  21606 & 24.61 $\pm$ 0.14\footnotemark[a] & 21602 & 21609 & 360  \\
Gemini North & $i'$  & 5985  & 23.46 $\pm$ 0.14 & 5982 & 5989 & 360 \\
Gemini North & $g'$ & 6002 & 24.14 $\pm$ 0.09 & 5998  & 6007  & 480 \\
\hline 
  \hline							
 \end{tabular}	
\footnotetext[a]{host galaxy dominated}							

 \medskip							
\end{table}							

}

\end{document}